\documentclass[aps,prl,longbibliography,twocolumn,superscriptaddress]{revtex4-2}
\usepackage{amsmath,amssymb,bm,graphicx,color,gensymb,bbold,hyperref,keyval,url,latexsym}
\usepackage[dvipsnames,svgnames,table]{xcolor}
\usepackage{enumitem}
\usepackage{ulem}

\hypersetup{pdfstartview={FitH}, colorlinks=true, linkcolor=NavyBlue, citecolor=Maroon, filecolor=NavyBlue, urlcolor=NavyBlue}

\begin{document}

\title{Spinless Charged Excitation at the Interface between a Conventional Topological Insulator and a Topological Mott Insulator}

\author{Cesar A. Gallegos}
\affiliation{Department of Physics and Astronomy, University of California, Irvine, California 92697, USA}
\author{Andrew J. Millis}
\affiliation{Center for Computational Quantum Physics, Flatiron Institute, New York, New York 10010, USA}
\affiliation{Department of Physics, Columbia University, New York, NY 10027, USA}
\author{Steven R. White}
\affiliation{Department of Physics and Astronomy, University of California, Irvine, California 92697, USA}
\date{\today}
\begin{abstract}
We investigate the interface separating two topologically distinct insulating phases of matter using extensive density-matrix renormalization group calculations to study the triangular-lattice Hofstadter-Hubbard model with a spatially varying interaction strength, chosen to realize both integer quantum Hall and chiral spin liquid states in different spatial regions.
We find that the integer quantum Hall-chiral spin liquid interface hosts a spinless charged excitation that is bound to the interface. 
This mode at the interface is identified through charge and spin pumping, and by direct calculations of low-lying excited states.  
We also characterize bulk excitations in both phases, finding evidence for fractionalization in the chiral spin liquid and for spin-triplet exciton formation in the integer quantum Hall phase.
\end{abstract}
\maketitle
{\it Introduction.}---%
Boundaries of  gapped topological phases often support gapless propagating edge modes, with properties determined by the change in wave function topology across the boundary~\cite{Halperin82,Wen1991,Qi06}. The most widely discussed is an ``edge", an interface separating a material, which may host a topologically nontrivial electronic state, from the vacuum, which is topologically trivial.
In this paper, we ask what happens at an interface separating two distinct topological phases. Such an interface is constrained by both bulks and does not need to reproduce the edge spectrum of either phase separately~\cite{Bais2009, Crepel2019, Zhu2020, Bollmann2024, Wagner2024, Pangburn2025, Wang2026}. 

A particularly interesting example is the interface between a topological insulator (TI) and a topological Mott insulator (TMI). The TI supports  gapless edge states carrying both charge and spin~\cite{Qi06}. In the TMI, strong correlations can fractionalize the electron, producing a neutral spinon edge mode that carries spin but no charge~\cite{Savary2016QSL}. At the TI--TMI interface, it has been proposed that the edge reconstructs in such a way that the spin sector is gapped while the charge sector remains gapless, so that the interface hosts a propagating spinless charged mode~\cite{Wagner2023, Wagner2024}.

In this Letter, we study a realization of the TI-TMI interface in the triangular-lattice Hofstadter-Hubbard model, which realizes~\cite{Knap2024, Divic2026, us2026} both an integer quantum Hall phase (IQH), a kind of TI state, and a chiral spin liquid (CSL), a TMI phase~\cite{Kalmeyer87}. 
In this model, the edge mode of the IQH phase carries both charge and spin, whereas the CSL edge supports only a neutral spinon mode. 
By making the Hubbard interaction spatially dependent, we construct systems in which one region is in the IQH phase and the other is in the CSL phase, with an interface between them.
To study this system we use  density matrix renormalization group (DMRG) calculations on finite-radius cylinders, allowing us to study both the topological response to threaded fluxes and the low-lying excited states localized near the IQH--CSL interface in the fully interacting lattice model.

We consider long but finite cylinders divided into two regions, one with a small value of the Hubbard interaction, such that this region of the system is in the IQH phase, and one with a large value, such that the corresponding region of the system is in the CSL phase. The system thus has an internal interface defined as the point at which the Hubbard interaction changes from small to large, and two external interfaces: the IQH-vacuum interface and the CSL-vacuum interface. 

Our main finding is that this IQH--CSL interface hosts a spinless charged excitation, and that this is the only gapless excitation strongly bound to the interface.
We demonstrate this in two complementary ways. First, flux insertion shows that charge that is pumped from the IQH edge halts at the IQH--CSL interface, while spin that is pumped from the IQH-vacuum interface passes through the internal interface, appearing at the CSL-vacuum interface.
Second, direct calculations of low-lying excited states show that a single-particle excitation near the interface separates into a charged component tightly bound to the interface and an unbound spinon.
Furthermore, we characterize bulk excitations in both phases. 
In the CSL phase, we find evidence for fractionalization, with single-particle excitations exhibiting spin-charge separation and charge-neutral spin-triplet excitations fractionalizing into two spatially separated spinons.
In the IQH phase, we find evidence for electron-hole binding in the spin-triplet channel, leading to exciton formation corresponding to low energy but gapped magnons.

\begin{figure}[t!]
\includegraphics[width=\linewidth]{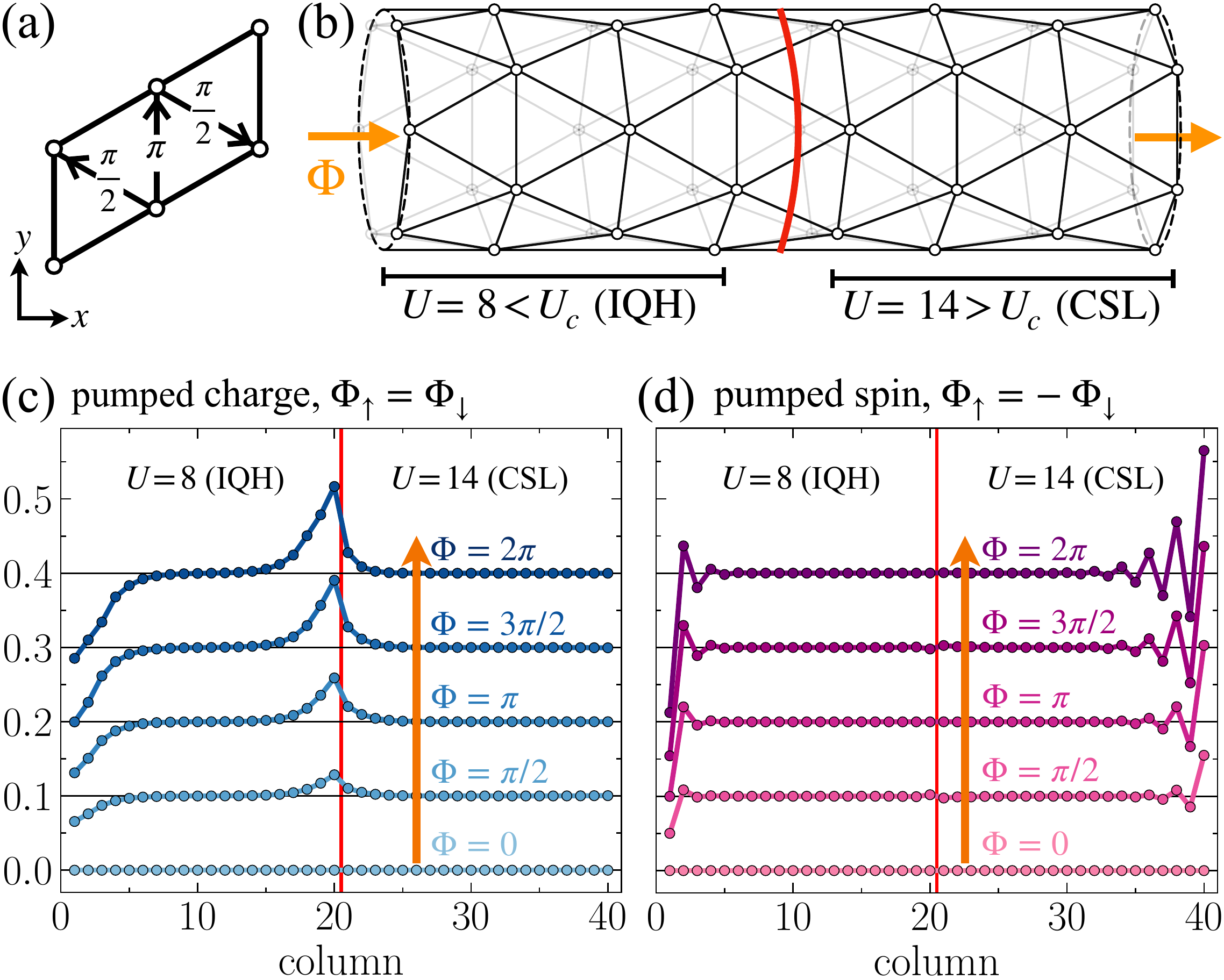}
\vskip -0.3cm
\caption{(a) Magnetic unit cell of the triangular lattice with flux $\pi/2$ per triangle. 
Nonzero Peierls hopping phases are shown.
(b) Representative IQH--CSL YC6 cylinder under flux insertion $\Phi$. 
Change in position dependent charge density  $n_{\Phi}\!-\!n_{\Phi=0}$ (c) and spin density $S^z_{\Phi}\!-\!S^z_{\Phi=0}$ (d) caused by threading the same and opposite fluxes for the two spin species, respectively, in $40\times6$ cylinders.
The charge and spin pumped for different fluxes are shifted by $0.1$ for clarity. The vertical red line in (b)-(d) indicates the IQH--CSL interface, and the orange large arrows indicate the pumping progression.}
\label{Fig:Pump}
\vskip -0.6cm
\end{figure}

{\it  Model and Method.}---%
The triangular-lattice
Hofstadter Hubbard model is defined by the Hamiltonian
\noindent
\vskip -0.35cm
\begin{equation}
H\! =\! -\sum_{\sigma, \langle ij\rangle} (t_{ij} c^\dagger_{i\sigma} c_{j\sigma}\!+\!{\rm h.c.}) \!-\!\sum_{\sigma,i}\varepsilon_i n_{i\sigma} \!+\!U \sum_i n_{i\uparrow}n_{i\downarrow},
\label{eq:HHmodel}
\end{equation}
\vskip -0.2cm
\noindent
where $i,j$ label sites on a triangular lattice, $\sigma\!=\, \uparrow,\downarrow$ denotes spin, $\langle ij\rangle$ correspond to the nearest-neighbor bonds, $t_{ij}$ are complex hoppings with phases encoding a flux of $\pi/2$ per triangle, see Fig.~\ref{Fig:Pump}(a), and $\varepsilon_i$ and $U$ are the on-site energy and Hubbard interaction, respectively.

In this Letter, we focus on the half-filled ground state, set the Zeeman coupling to zero, and use $|t_{ij}|=1$ as the energy scale throughout. 
For the subsequent discussion of charge and spin pumping we also consider threading the cylinder with a flux $\Phi$ which may be the same  or opposite in sign for the two spin species. 
The flux is implemented as an additional Peierls phase arising from a uniform vector potential in the $y$ direction, which does not break $y$ translational symmetry~\footnote{The vertical bonds pick up an additional phase of $\Phi/L_y$, while the extra phase of the diagonal bonds is smaller by a factor of two (i.e. $\sin \pi/6$); see Figs.~\ref{Fig:Pump}(a) and (b).}.

The noninteracting band structure of the model (Eq.~\eqref{eq:HHmodel} at $U\!=\!0$) hosts Chern bands separated by a gap around zero energy, so that at half-filling and for $U\!<\!U_c\!\approx\!11$, the ground state is an IQH state with Hall number $\nu=1$ for each spin.  
At very large $U$, standard strong coupling arguments map the model onto the triangular lattice antiferromagnetic Heisenberg model, whose ground state is believed to be a topologically trivial magnetically ordered state. 
Recent computational work suggests that the small $U$ and very large $U$ phases are separated by a topologically distinct intermediate-$U$ phase that has been identified as a CSL~\cite{Knap2024, Divic2026, us2026}, providing a natural setting to analyze excitations at an interface between two topologically distinct insulators.

We use DMRG calculations on finite cylinders that are periodic in the transverse direction and open in the longitudinal direction. 
We studied cylinders with transverse periodicity $3$, $5$ and $6$, labeled as YC$3$, YC$5$, YC$6$ consistent with the notation in Refs.~\cite{Knap2024, Divic2026, us2026}.
For the interface calculations, we take the Hubbard interaction to be spatially dependent, choosing the left part of the cylinder to be in the IQH phase, with $U=8$, and the right part to be in the CSL phase, with $U=14$ with the interaction changing across one bond; see Fig.~\ref{Fig:Pump}(b). 
The on-site energy is chosen as $\varepsilon_i= \!U_i/2$, and the chemical potential is set to zero so that all portions of the cylinder remain at one electron per site and particle-hole symmetry is preserved. To efficiently obtain the ground state of the IQH-CSL cylinders it was found to be helpful to start from the exact $U=0$ single-particle IQH state~\cite{Fishman2015GaussianMPS} and then increase $U$ every other sweep. 
The states are characterized by the column-averaged charge density $n(x)\!=\!\frac{1}{L_y}\sum_{y} \langle n_{x,y}\rangle$ and magnetization
$S^z(x)\!=\!\frac{1}{L_y}\sum_{y} \langle S^z_{x,y}\rangle$, as well as the excitation spectra and charge and spin pumping properties.

{\it  IQH--CSL interface: charge and spin pump.}---%
We first consider the charge and spin pumping properties, and in particular the response of the system to fluxes $\Phi$ threaded along the cylinder axis and adiabatically increased from $0$ to $2\pi$~\cite{Zaletel2014}.
For charge pumping we take the flux to have the same sign and magnitude for both spin species; for spin pumping the flux is equal in magnitude but opposite in sign for spin up and spin down.
In the IQH state, the charge pump transfers two electrons, one for each spin species, as the flux increases from $0$ to $2\pi$, while the spin pump transfers a net spin of 1. 
In the CSL, the system does not pump charge, consistent with a Mott-insulating state, but the spin pump still transfers one unit of spin~\cite{Knap2024, us2026}, as expected from the relation between the $\nu=1/2$ fractional quantum Hall state and the CSL~\cite{Kalmeyer87}.

Figures~\ref{Fig:Pump}(c) and \ref{Fig:Pump}(d) show the change in the charge density, $n_{\Phi}-n_{\Phi=0}$, and magnetization, $S^z_{\Phi}-S^z_{\Phi=0}$, profiles in the YC6 cylinder under flux insertion with the same and opposite fluxes for the two spin species, respectively.
The results were obtained by starting from the converged ground-state wavefunction at $\Phi\!=\!0$ and increasing the flux in steps of $\pi/4$. At each step, two DMRG sweeps were performed with bond dimension up to $\chi\!=\!6000$, yielding a truncation error of order $\mathcal{O}(10^{-5})$ in the $40\!\times\!6$ YC cylinder. In the calculations shown, we find no sudden change in the energy, indicating that the state follows a continuous adiabatic path.

We see from Fig.~\ref{Fig:Pump}(c) that under charge pumping conditions, charge is removed from the IQH-vacuum interface and accumulates at the IQH--CSL interface, consistent with the idea that because the CSL is a Mott insulator, it does not support a charge pump. The accumulation of pumped charge at the IQH--CSL interface suggests that the IQH-CSL interface can support a charge mode.

Fig.~\ref{Fig:Pump}(d) shows that under spin pumping conditions spin density is removed from the IQH-vacuum interface and transferred to the CSL-vacuum interface, with negligible spin density appearing at the IQH-CSL interface. 
Instead, spin is transported through both phases, from one physical edge to the other, indicating that the interface does not trap spin in the pumping process. 
Together, the charge and spin pumps provide strong evidence that the IQH--CSL interface supports a spinless charged mode.

\begin{figure}[t!]
\includegraphics[width=\linewidth]{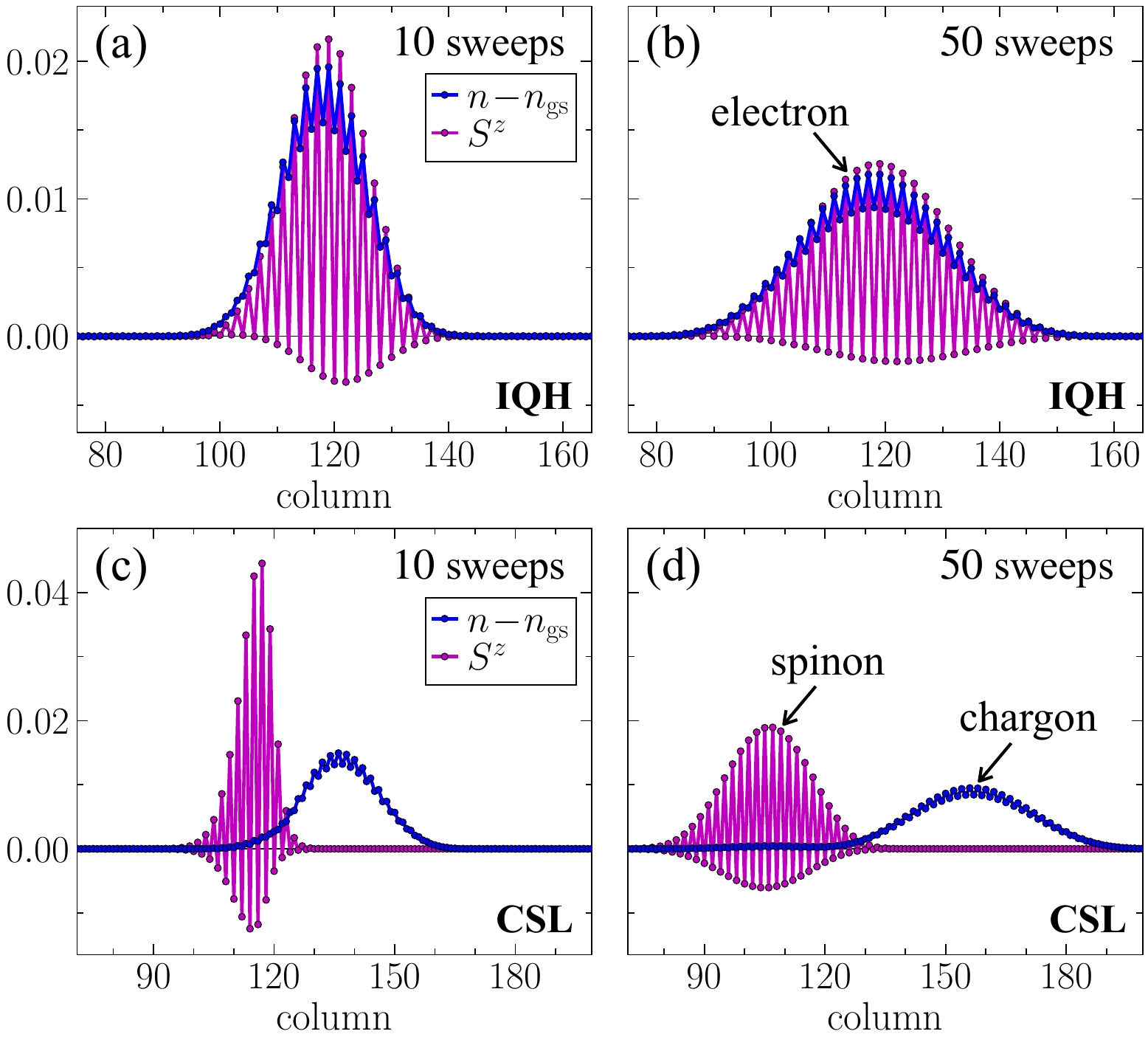}
\vskip -0.3cm
\caption{Single-particle bulk excitations in the [(a), (b)] IQH and [(c), (d)] CSL phases at $U\!=\!8$ and $U\!=\!14$, respectively, in a $240\times3$ cylinder keeping a bond dimension $\chi\!=\!800$.
The solid blue and purple lines show the deviation of the charge density from the ground state, $n-n_{\rm gs}$, and the local magnetization, $S^z$, respectively, for different sweeps. The electron-like excitation in the IQH and its spin-charge separation into a spinon and chargon in the CSL phase are indicated.}
\label{Fig:1p}
\vskip -0.6cm
\end{figure}

{\it Fractionalization in bulk phases.}---%
We now investigate excitations in the bulk phases by creating a trial state that  corresponds to starting from one of the two bulk ground states and either  adding or removing an electron from it or creating a spin flip (magnon), and then using DMRG to relax the trial state to obtain an approximation of an eigenstate.  
Our trial states differ from the ground state by a spin and perhaps a charge quantum number. 
The DMRG procedure optimizes this state, driving it towards the lowest lying state consistent with the quantum numbers and bond dimension (see End Matter). 
The initial trial states are chosen to be of the form $|\psi\rangle \!=\! \sum_{x,y} f(x) \hat{A}_{(x,y)}|\psi_{\rm gs}\rangle$, where $|\psi_{\rm gs}\rangle$ is the ground state, $f(x)$ is a Gaussian envelope with a width of $6$ columns centered at a particular point, and $\hat{A}_{(x,y)}\!=\!c_{{(x,y)},\uparrow}$ and $c^\dagger_{(x,y),\uparrow}$ for the single-particle excitations and $\hat{A}_{(x,y)}\!=\!S^+_{(x,y)}$ for the spin-triplet excitations.
These calculations were performed on $240\!\times\!3$ YC cylinders with uniform interaction, using $U\!=\!8$ and $U\!=\!14$ for the IQH and CSL phases, respectively.
We typically performed 50 sweeps and kept up to $\chi\!=\!800$ states, yielding a truncation error of ${\cal O}(10^{-6})$.
In most cases the form of the excitation itself is established after a small number of DMRG sweeps at fixed bond dimension. 

At fixed bond dimension, additional sweeps slowly relax the energy of the system, while retaining the spatial form. Increasing the bond dimension broadens the optimized charge density and magnetization profiles of the excited states in the IQH and CSL phases. We expect that in the limit of infinite bond dimension and infinite number of DMRG sweeps the spatial profile  would relax to a plane-wave excitation. 
Therefore, although the states shown below are not fully converged with respect to bond dimension, they already display the structure of the relevant excitations; see End Matter for further discussion.

Figure~\ref{Fig:1p} shows the single-particle excitation obtained by adding an electron to the ground state in the bulk of the cylinder, both in the IQH  (Figs.~\ref{Fig:1p}(a) and (b)) and CSL  (Figs.~\ref{Fig:1p}(c) and (d)), for a smaller and larger number of DMRG sweeps, at a fixed bond dimension. 
The added electron creates a state of zero average momentum.
In the IQH phase, the single-particle excitation behaves as expected for an electron, with the charge and spin densities remaining spatially bound together. 
In the IQH state, the resulting electron wave function spreads symmetrically, as it evolves into a progressively lower energy superposition of positive and negative momenta, consistent with the vanishing average momentum of the created state. 
After a moderate amount of sweeps $(\approx 70$ for $\chi=800$), the energy converges with an error of ${\cal O}(10^{-5})$.
Additional DMRG sweeps and larger bond dimension reveal a spreading of the excitation
without separating its charge and spin components; see Figs.~\ref{Fig:1p}(a) and (b).

In the CSL phase, Figs.~\ref{Fig:1p}(c) and (d), the single-particle excitation rapidly (i.e. in only a few DMRG sweeps) separates into two peaks, one containing the added charge (``chargon") and one the added spin (``spinon"), consistent with the electron fractionalization expected in quantum spin liquids~\cite{Savary2016QSL, Knolle2019QSL}. 
As the DMRG sweeps continue and the bond dimension $\chi$ is increased, the charge and spin density profiles broaden and gradually separate further from one another, decreasing their kinetic energy. 
Whether the spinon is to the left or right of the chargon depends on initial sweeping details that are not at present understood.

\begin{figure}[t!]
\includegraphics[width=\linewidth]{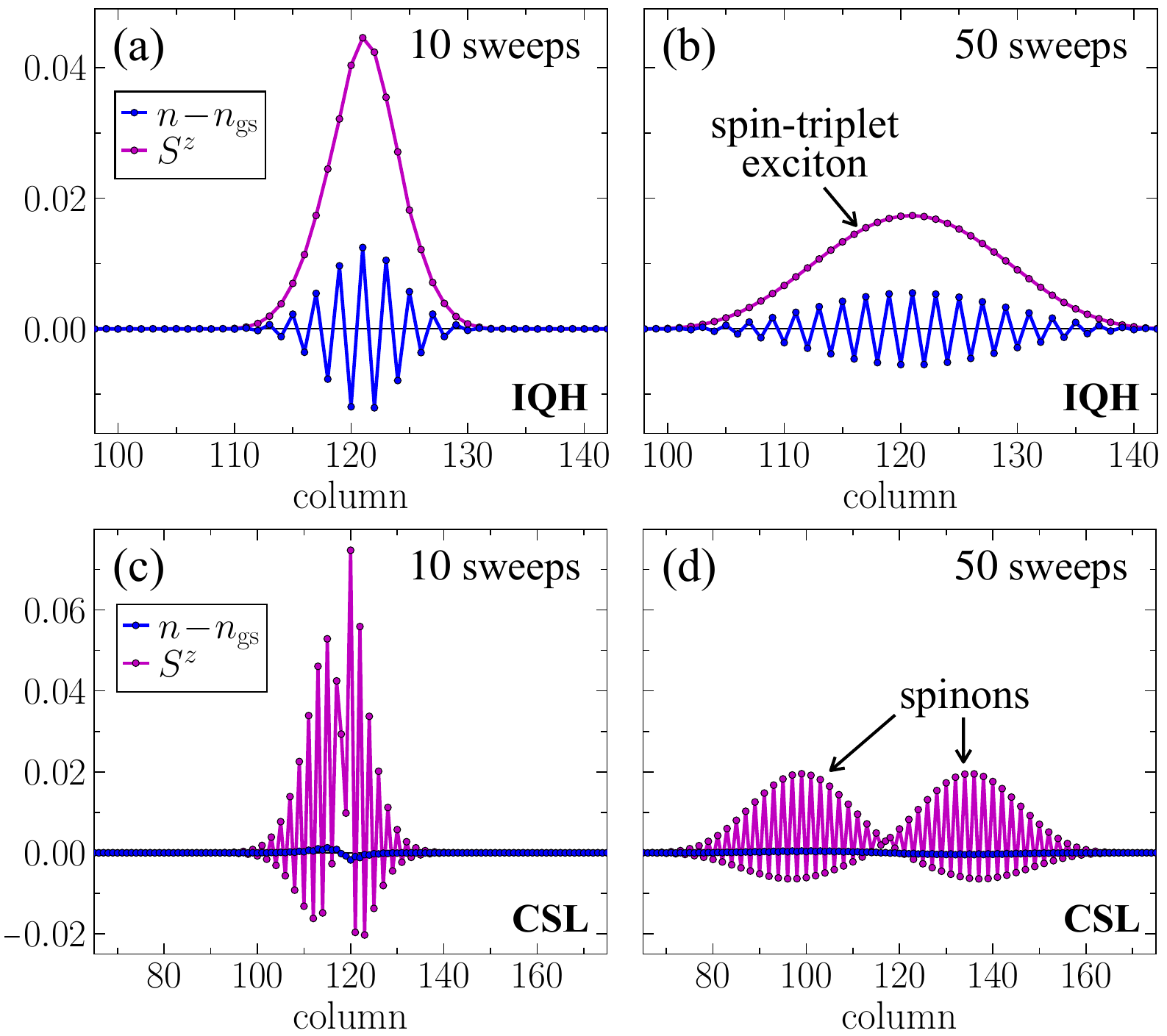}
\vskip -0.3cm
\caption{Same as Fig.~\ref{Fig:1p}, but for the charge-neutral spin-triplet bulk excitation.
The spin-triplet exciton in the IQH phase and the fractionalization into two spatially separated spinons in the CSL phase are indicated.}
\label{Fig:S+}
\vskip -0.5cm
\end{figure}

We next consider a charge-neutral spin-triplet excitation obtained by creating a magnon by adding a spin up and removing a spin down electron in the bulk of the cylinder. 
As above, Fig.~\ref{Fig:S+} shows the relaxation of the resulting state during the DMRG optimization for a smaller and larger number of sweeps at a fixed bond dimension, comparing the IQH phase, Figs.~\ref{Fig:S+}(a) and (b), with the CSL phase, Figs.~\ref{Fig:S+}(c) and (d).  
Figs.~\ref{Fig:S+}(a) and (b) show that in the IQH phase the two components of the excitation remain bound together, corresponding in effect to a spin one exciton in the gap of the IQH state~\cite{Halperin1984}. Figs.~\ref{Fig:S+}(c) and (d) show that unlike in the IQH phase, the charge-neutral spin-triplet excitation in the CSL phase separates into two components, each carrying $S^z\!=\!1/2$, which broaden and separate further as the DMRG optimization proceeds and the bond dimension $\chi$ is increased, as expected in the fractionalized CSL phase~\cite{Savary2016QSL, Knolle2019QSL}.
We also note that the spinons are accompanied by a small charge-density deviation of ${\cal O}(10^{-4})$ during the first few DMRG sweeps, which becomes negligible once the spinons are separated by a few columns, whereas the spin 1 excitation in IQH phase is accompanied by a clear oscillating charge density modulation.

{\it Spin-triplet exciton in the IQH phase.}---%
The charge-neutral spin-triplet excitation in the IQH phase, shown in Fig.~\ref{Fig:S+}(b), is naturally interpreted as an electron-hole pair in the spin-triplet channel~\cite{Halperin1984}.
We define its binding energy as
\noindent
\vskip -0.65cm
\begin{align}
E_{\rm bind}^{\rm T}=E_{e+h}-E_{\rm T},
\label{eq:Ebind_triplet}
\end{align}
\vskip -0.15cm
\noindent
where $E_{e+h}$ is the energy of a well-separated electron-hole pair and $E_{\rm T}$ is the lowest energy in the charge-neutral spin-triplet sector.
With this convention, a positive $E_{\rm bind}^{\rm T}$ indicates a bound spin-triplet exciton.

\begin{figure}[t!]
\includegraphics[width=\linewidth]{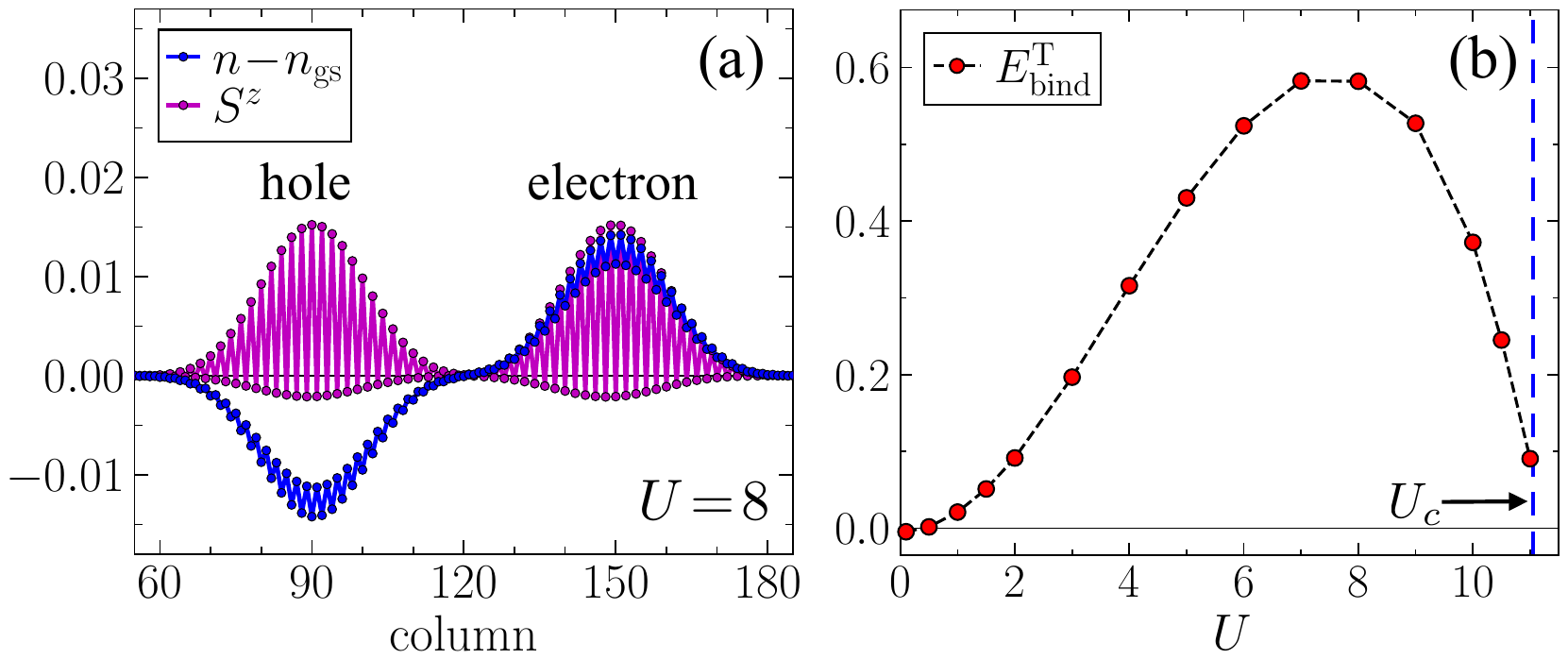}
\vskip -0.3cm
\caption{(a) A separated electron-hole pair in the IQH phase at $U\!=\!8$ in a $240\times3$ cylinder. 
The solid blue and purple lines show the deviation of the charge density from the ground state, $n\!-\!n_{\rm gs}$, and the local magnetization, $S^z$, respectively.
(b) The spin-triplet electron-hole binding energy from Eq.~\eqref{eq:Ebind_triplet} as a function of $U$ in the IQH phase.
The blue dashed line indicates the IQH--CSL transition point $U_c$.}
\label{Fig:ExcitonTriplet}
\vskip -0.6cm
\end{figure}

Figure~\ref{Fig:ExcitonTriplet}(a) shows a state in the spin-triplet $(S^z\!=\!1)$ sector obtained at $\chi\!=\!800$ after initializing DMRG from an electron-hole pair separated by $60$ columns in the IQH ground state.
This state provides a numerical estimate of $E_{e+h}$ in Eq.~\eqref{eq:Ebind_triplet}.
When the electron and hole are initialized closer to one another, or when additional DMRG sweeps are performed at larger bond dimension, the electron and hole densities reorganize to form the spin-triplet exciton shown in Fig.~\ref{Fig:S+}(b).

The corresponding calculation in the CSL does not yield an electron-hole bound state.
Whether the electron and hole are initialized close to one another or spatially separated, spin charge separation occurs:
during the DMRG sweeps, the spinless charge excitations move toward the edges of the cylinder, while the spinons remain in the bulk and form a pattern similar to that shown in Fig.~\ref{Fig:S+}(d).

Figure~\ref{Fig:ExcitonTriplet}(b) shows the spin-triplet electron-hole binding energy from Eq.~\eqref{eq:Ebind_triplet} as a function of $U$ in the IQH phase.
The small-$U$ quadratic behavior $E_{\rm bind}^{\rm T}\!\sim \!U^2$ is consistent with a leading attraction in the triplet channel arising from a second-order process in perturbation theory.
The binding energy is maximal at $U\!\approx \!7.5$.
Near the critical interaction $U_c$, where the IQH--CSL transition occurs, our calculations on the YC3 cylinder suggest that the binding energy does not vanish, although we cannot assert that this holds in wider cylinders.

{\it  IQH--CSL interface: excited states.}---%
We now investigate interface states in more detail by creating  trial states similar to those discussed above in the bulk case but now in the vicinity of the interface then using DMRG to relax the trial state to obtain an approximation of an eigenstate. 
In these calculations we keep a bond dimension of up to $\chi\!=\!800$ in the $240\!\times\!3$ and $\chi\!=\!1500$ in the $80\!\times\!5$ YC cylinders, yielding truncation errors of ${\cal O}(10^{-6})$ and ${\cal O}(10^{-5})$ respectively.

For odd-circumference cylinders, the IQH ground state hosts opposite edge charge excesses of $1/2$, and threading a $\pi$-flux flips the left and right edge charges in the IQH phase~\cite{us2026}.
In the YC3 cylinders, the IQH--CSL interface has an excess charge of $1/2$, and we thread a $\pi$-flux to obtain a charge of $-1/2$ at the interface. 
This way, adding an electron yields the smallest net charge at the IQH--CSL interface, $-1/2+1\!=\!1/2$, thus minimizing the finite-width repulsion between charges at the interface; see End Matter.
In the YC5 cylinders, the interface has a charge deficit of $1/2$, so the additional flux is not needed.

\begin{figure}[t!]
\includegraphics[width=\linewidth]{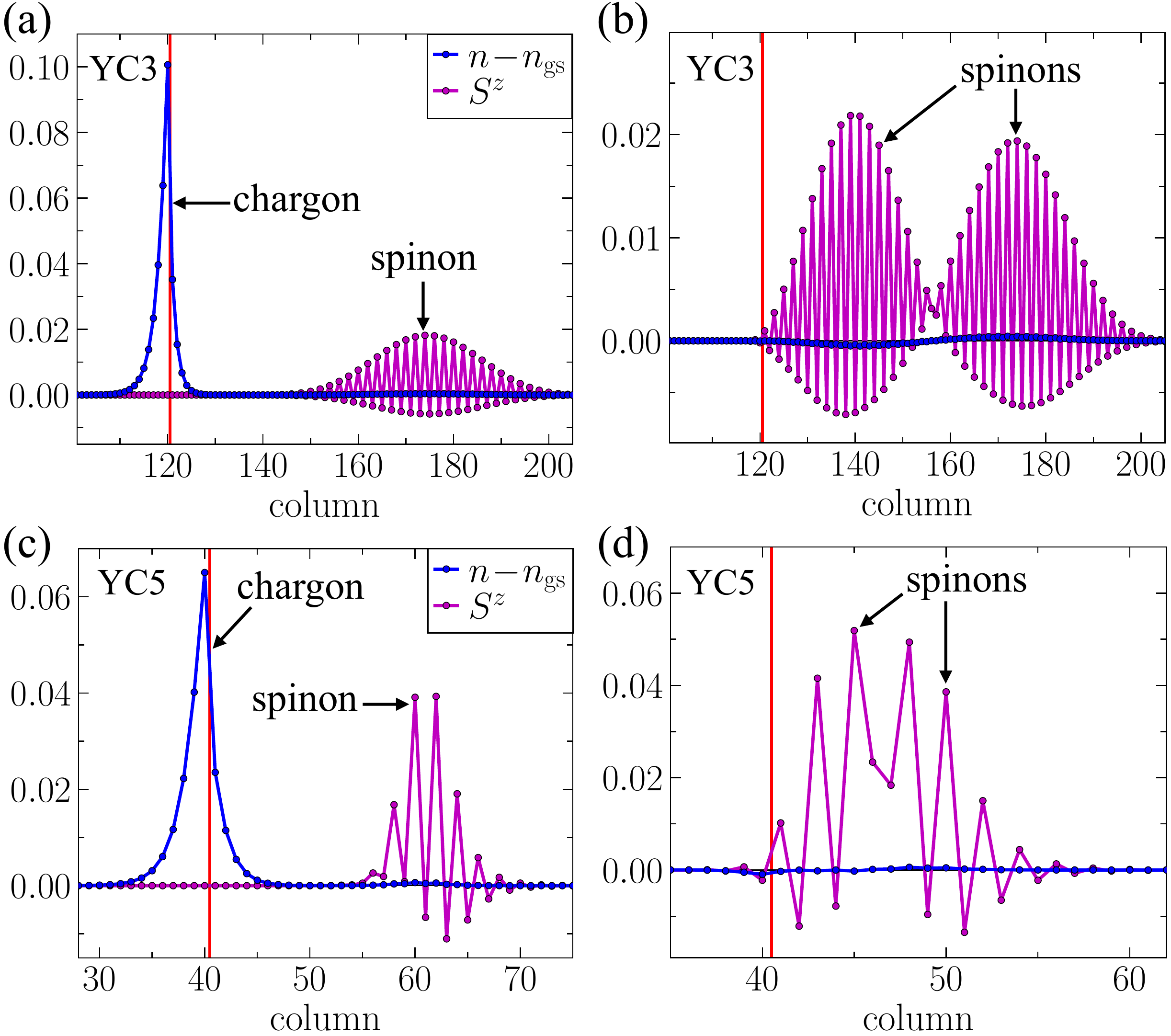}
\vskip -0.3cm
\caption{Excited states at the IQH--CSL interface.
The [(a), (c)] single-particle and [(b), (d)] charge-neutral spin-triplet excitation for [(a), (b)] YC3 (length $240$) and [(c), (d)] YC5 (length $80$) cylinders. 
The solid red line marks the IQH--CSL interface.
The solid blue and purple lines show the deviation of the charge density from the ground state, $n\!-\!n_{\rm gs}$, and the local magnetization, $S^z$, respectively.
The spinless charged excitations, chargon and holon, and the fractionalized spin excitation, spinon, are indicated.}
\label{Fig:Interface1pSp}
\vskip -0.6cm
\end{figure}

Figs.~\ref{Fig:Interface1pSp}(a) and (c) show the charge and spin densities resulting from relaxing an initial added electron state. 
The single-particle excitation at the IQH--CSL interface reveals a nontrivial fractionalization process. When an electron is created near the interface, the chargon remains strongly localized at the interface, while the spinon is expelled into the CSL region. 
The subsequent motion of the spinon away from the interface is very slow, requiring many DMRG sweeps to resolve, suggesting that the spinon-interface interaction is very local, governed by the tail of the wavepacket touching the interface. 

We verify this interpretation using two complementary approaches. 
First, starting from a state in which the spinon density remains closer to the internal interface than to the free edge after many sweeps, we apply a force to the spinon by adding a magnetic field which increases linearly inside the CSL region and moves the spinon away from the interface and into the center of the CSL within $\approx 5$ sweeps.
We then turn off the field and continue sweeping at the same bond dimension, so that the change in energy is associated with the displacement of the spinon. 
On YC3 and YC5 this gives an energy gain of ${\cal O}(10^{-4})$ and ${\cal O}(10^{-3})$, respectively, suggesting effective repulsion between the interface and the spinon. 
Second, we create the electron directly at the center of the CSL. The excitation again separates; the charge moves to the interface while the spin stays in the CSL. The two calculations give states with the same energy, within our accuracy, and spatial structure. Thus, the lowest single-particle excitation at the IQH--CSL interface is consistent with a spinless charged edge mode.

The charge-neutral spin-triplet excitation in the YC3 cylinder, Fig.~\ref{Fig:Interface1pSp}(b), shows a clear separation of the $S^z\!=\!1$ excitation into a pair of spatially separated spinons, each carrying $S^z\!=\!1/2$ and located in the general vicinity of the interface on the CSL side.
The separation of the spinons in the wider YC5 cylinder is less pronounced, but a similar pattern remains visible.
In addition, the two spinons in YC3 are accompanied by a small ${\cal O}(10^{-4})$ charge-density deviation, which is suppressed by an order of magnitude in the wider YC5 cylinder.
In all cases, both spinons remain closer to the IQH--CSL interface, but are slowly displaced toward the CSL bulk as the sweeps continue. This behavior is consistent with the absence of binding between the spinons and the interface inferred above.

Overall, the excited states at the IQH--CSL interface provide strong evidence for a spinless charged mode tightly bound to the interface.
The spin component, by contrast, lies on the CSL side and is repelled from the interface.

{\it Summary.}---%
We have studied the excitations at the interface between an integer quantum Hall (IQH) phase and a chiral spin liquid (CSL) in the triangular-lattice Hofstadter-Hubbard model.
Using DMRG on cylinders with a spatially dependent Hubbard interaction, we have shown that the IQH--CSL interface supports a spinless charged excitation, as suggested in Ref.~\cite{Wagner2024}.
The flux-insertion response provides one characterization of this mode.
The charge pump transfers charge from the IQH edge to the IQH--CSL interface, while the spin pump transfers spin through the cylinder.
Direct calculations of excited states give a consistent picture, in which the charge component of a single-particle excitation remains tightly bound to the interface, while the spin degree of freedom appears as a spinon repelled from the interface, implying an exotic edge state that supports charge $e$ chargon excitations but not spinon excitations. 

The IQH–CSL interface can be understood, following Ref.~\cite{Ebisu2020} by noting that the spin-degenerate $\nu\!=\!2$ IQH edge decomposes into a $U(1)_1$ charge sector and an $SU(2)_1$ spin sector, while the CSL edge can be described by a counterpropagating  $SU(2)_1$ mode. 
For our  model, the spin coupling  is  antiferromagnetic; this marginally relevant interaction gaps the spin sectors leaving a gapless charge sector whose minimal excitation has charge $e$ and spin zero, consistent with our numerical results~\cite{footnote}.

We have also analyzed bulk excitations in the CSL and IQH phases.
In the CSL phase, we find evidence for fractionalization, with single-particle excitations separating into spin and charge components and charge-neutral spin-triplet excitations splitting into two spatially separated spinons.
In the IQH phase, single-particle excitations remain electron-like, with charge and spin bound together, while charge-neutral spin excitations form a spin-triplet exciton.
The binding energy of this spin-triplet exciton scales quadratically with the Hubbard interaction at small $U$ but becomes small near the critical $U$ for the IQH-CSL transition. The small but apparently non-zero binding energy near the transition may be relevant in the context of phases occurring near the transition point~\cite{Divic2025, kuhlenkamp2025, Chen2026, Niu2026}, and warrants further investigation.

\begin{acknowledgments}
{\it Acknowledgments.}---%
We thank S. Divic, T. Soejima, U. Schollwoeck, D. Guerci, E. König and K. Yang for helpful discussions.
CAG was supported by the UCI-LANL-SoCal Hub Graduate Fellowship program and by the Eddleman Quantum Institute at UCI. 
SRW was supported by the U.S. NSF under Grant DMR-2412638. 
The work of AJM was supported in part by the National Science Foundation through the Columbia University Materials Research Science and Engineering Center (MRSEC) under NSF Award No. DMR-2614607.
The calculations were performed using the ITensor library~\cite{itensor}. 
The Flatiron Institute is a division of the Simons Foundation.
\end{acknowledgments}

\bibliography{IQH-CSL_Interface}

\begin{thebibliography}{28}%
\makeatletter
\providecommand \@ifxundefined [1]{%
 \@ifx{#1\undefined}
}%
\providecommand \@ifnum [1]{%
 \ifnum #1\expandafter \@firstoftwo
 \else \expandafter \@secondoftwo
 \fi
}%
\providecommand \@ifx [1]{%
 \ifx #1\expandafter \@firstoftwo
 \else \expandafter \@secondoftwo
 \fi
}%
\providecommand \natexlab [1]{#1}%
\providecommand \enquote  [1]{``#1''}%
\providecommand \bibnamefont  [1]{#1}%
\providecommand \bibfnamefont [1]{#1}%
\providecommand \citenamefont [1]{#1}%
\providecommand \href@noop [0]{\@secondoftwo}%
\providecommand \href [0]{\begingroup \@sanitize@url \@href}%
\providecommand \@href[1]{\@@startlink{#1}\@@href}%
\providecommand \@@href[1]{\endgroup#1\@@endlink}%
\providecommand \@sanitize@url [0]{\catcode `\\12\catcode `\$12\catcode `\&12\catcode `\#12\catcode `\^12\catcode `\_12\catcode `\%12\relax}%
\providecommand \@@startlink[1]{}%
\providecommand \@@endlink[0]{}%
\providecommand \url  [0]{\begingroup\@sanitize@url \@url }%
\providecommand \@url [1]{\endgroup\@href {#1}{\urlprefix }}%
\providecommand \urlprefix  [0]{URL }%
\providecommand \Eprint [0]{\href }%
\providecommand \doibase [0]{https://doi.org/}%
\providecommand \selectlanguage [0]{\@gobble}%
\providecommand \bibinfo  [0]{\@secondoftwo}%
\providecommand \bibfield  [0]{\@secondoftwo}%
\providecommand \translation [1]{[#1]}%
\providecommand \BibitemOpen [0]{}%
\providecommand \bibitemStop [0]{}%
\providecommand \bibitemNoStop [0]{.\EOS\space}%
\providecommand \EOS [0]{\spacefactor3000\relax}%
\providecommand \BibitemShut  [1]{\csname bibitem#1\endcsname}%
\let\auto@bib@innerbib\@empty
\bibitem [{\citenamefont {Halperin}(1982)}]{Halperin82}%
  \BibitemOpen
  \bibfield  {author} {\bibinfo {author} {\bibfnamefont {B.~I.}\ \bibnamefont {Halperin}},\ }\bibfield  {title} {\bibinfo {title} {{Quantized Hall conductance, current-carrying edge states, and the existence of extended states in a two-dimensional disordered potential}},\ }\href {https://doi.org/10.1103/PhysRevB.25.2185} {\bibfield  {journal} {\bibinfo  {journal} {Phys. Rev. B}\ }\textbf {\bibinfo {volume} {25}},\ \bibinfo {pages} {2185} (\bibinfo {year} {1982})}\BibitemShut {NoStop}%
\bibitem [{\citenamefont {Wen}(1991)}]{Wen1991}%
  \BibitemOpen
  \bibfield  {author} {\bibinfo {author} {\bibfnamefont {X.~G.}\ \bibnamefont {Wen}},\ }\bibfield  {title} {\bibinfo {title} {{Gapless boundary excitations in the quantum Hall states and in the chiral spin states}},\ }\href {https://doi.org/10.1103/PhysRevB.43.11025} {\bibfield  {journal} {\bibinfo  {journal} {Phys. Rev. B}\ }\textbf {\bibinfo {volume} {43}},\ \bibinfo {pages} {11025} (\bibinfo {year} {1991})}\BibitemShut {NoStop}%
\bibitem [{\citenamefont {Qi}\ \emph {et~al.}(2006)\citenamefont {Qi}, \citenamefont {Wu},\ and\ \citenamefont {Zhang}}]{Qi06}%
  \BibitemOpen
  \bibfield  {author} {\bibinfo {author} {\bibfnamefont {X.-L.}\ \bibnamefont {Qi}}, \bibinfo {author} {\bibfnamefont {Y.-S.}\ \bibnamefont {Wu}},\ and\ \bibinfo {author} {\bibfnamefont {S.-C.}\ \bibnamefont {Zhang}},\ }\bibfield  {title} {\bibinfo {title} {{General theorem relating the bulk topological number to edge states in two-dimensional insulators}},\ }\href {https://doi.org/10.1103/PhysRevB.74.045125} {\bibfield  {journal} {\bibinfo  {journal} {Phys. Rev. B}\ }\textbf {\bibinfo {volume} {74}},\ \bibinfo {pages} {045125} (\bibinfo {year} {2006})}\BibitemShut {NoStop}%
\bibitem [{\citenamefont {Bais}\ \emph {et~al.}(2009)\citenamefont {Bais}, \citenamefont {Slingerland},\ and\ \citenamefont {Haaker}}]{Bais2009}%
  \BibitemOpen
  \bibfield  {author} {\bibinfo {author} {\bibfnamefont {F.~A.}\ \bibnamefont {Bais}}, \bibinfo {author} {\bibfnamefont {J.~K.}\ \bibnamefont {Slingerland}},\ and\ \bibinfo {author} {\bibfnamefont {S.~M.}\ \bibnamefont {Haaker}},\ }\bibfield  {title} {\bibinfo {title} {{Theory of Topological Edges and Domain Walls}},\ }\href {https://doi.org/10.1103/PhysRevLett.102.220403} {\bibfield  {journal} {\bibinfo  {journal} {Phys. Rev. Lett.}\ }\textbf {\bibinfo {volume} {102}},\ \bibinfo {pages} {220403} (\bibinfo {year} {2009})}\BibitemShut {NoStop}%
\bibitem [{\citenamefont {Crépel}\ \emph {et~al.}(2019)\citenamefont {Crépel}, \citenamefont {Claussen}, \citenamefont {Estienne},\ and\ \citenamefont {Regnault}}]{Crepel2019}%
  \BibitemOpen
  \bibfield  {author} {\bibinfo {author} {\bibfnamefont {V.}~\bibnamefont {Crépel}}, \bibinfo {author} {\bibfnamefont {N.}~\bibnamefont {Claussen}}, \bibinfo {author} {\bibfnamefont {B.}~\bibnamefont {Estienne}},\ and\ \bibinfo {author} {\bibfnamefont {N.}~\bibnamefont {Regnault}},\ }\bibfield  {title} {\bibinfo {title} {{Model states for a class of chiral topological order interfaces}},\ }\href {http://dx.doi.org/10.1038/s41467-019-09168-z} {\bibfield  {journal} {\bibinfo  {journal} {Nat Commun}\ }\textbf {\bibinfo {volume} {10}},\ \bibinfo {pages} {1861} (\bibinfo {year} {2019})}\BibitemShut {NoStop}%
\bibitem [{\citenamefont {Zhu}\ \emph {et~al.}(2020)\citenamefont {Zhu}, \citenamefont {Sheng},\ and\ \citenamefont {Yang}}]{Zhu2020}%
  \BibitemOpen
  \bibfield  {author} {\bibinfo {author} {\bibfnamefont {W.}~\bibnamefont {Zhu}}, \bibinfo {author} {\bibfnamefont {D.~N.}\ \bibnamefont {Sheng}},\ and\ \bibinfo {author} {\bibfnamefont {K.}~\bibnamefont {Yang}},\ }\bibfield  {title} {\bibinfo {title} {{Topological Interface between Pfaffian and Anti-Pfaffian Order in $\ensuremath{\nu}=5/2$ Quantum Hall Effect}},\ }\href {https://doi.org/10.1103/PhysRevLett.125.146802} {\bibfield  {journal} {\bibinfo  {journal} {Phys. Rev. Lett.}\ }\textbf {\bibinfo {volume} {125}},\ \bibinfo {pages} {146802} (\bibinfo {year} {2020})}\BibitemShut {NoStop}%
\bibitem [{\citenamefont {Bollmann}\ \emph {et~al.}(2024)\citenamefont {Bollmann}, \citenamefont {Setty}, \citenamefont {Seifert},\ and\ \citenamefont {K\"onig}}]{Bollmann2024}%
  \BibitemOpen
  \bibfield  {author} {\bibinfo {author} {\bibfnamefont {S.}~\bibnamefont {Bollmann}}, \bibinfo {author} {\bibfnamefont {C.}~\bibnamefont {Setty}}, \bibinfo {author} {\bibfnamefont {U.~F.~P.}\ \bibnamefont {Seifert}},\ and\ \bibinfo {author} {\bibfnamefont {E.~J.}\ \bibnamefont {K\"onig}},\ }\bibfield  {title} {\bibinfo {title} {{Topological Green's Function Zeros in an Exactly Solved Model and Beyond}},\ }\href {https://doi.org/10.1103/PhysRevLett.133.136504} {\bibfield  {journal} {\bibinfo  {journal} {Phys. Rev. Lett.}\ }\textbf {\bibinfo {volume} {133}},\ \bibinfo {pages} {136504} (\bibinfo {year} {2024})}\BibitemShut {NoStop}%
\bibitem [{\citenamefont {Wagner}\ \emph {et~al.}(2024)\citenamefont {Wagner}, \citenamefont {Guerci}, \citenamefont {Millis},\ and\ \citenamefont {Sangiovanni}}]{Wagner2024}%
  \BibitemOpen
  \bibfield  {author} {\bibinfo {author} {\bibfnamefont {N.}~\bibnamefont {Wagner}}, \bibinfo {author} {\bibfnamefont {D.}~\bibnamefont {Guerci}}, \bibinfo {author} {\bibfnamefont {A.~J.}\ \bibnamefont {Millis}},\ and\ \bibinfo {author} {\bibfnamefont {G.}~\bibnamefont {Sangiovanni}},\ }\bibfield  {title} {\bibinfo {title} {{Edge Zeros and Boundary Spinons in Topological Mott Insulators}},\ }\href {https://doi.org/10.1103/PhysRevLett.133.126504} {\bibfield  {journal} {\bibinfo  {journal} {Phys. Rev. Lett.}\ }\textbf {\bibinfo {volume} {133}},\ \bibinfo {pages} {126504} (\bibinfo {year} {2024})}\BibitemShut {NoStop}%
\bibitem [{\citenamefont {Pangburn}\ \emph {et~al.}(2025)\citenamefont {Pangburn}, \citenamefont {P\'epin},\ and\ \citenamefont {Banerjee}}]{Pangburn2025}%
  \BibitemOpen
  \bibfield  {author} {\bibinfo {author} {\bibfnamefont {E.}~\bibnamefont {Pangburn}}, \bibinfo {author} {\bibfnamefont {C.}~\bibnamefont {P\'epin}},\ and\ \bibinfo {author} {\bibfnamefont {A.}~\bibnamefont {Banerjee}},\ }\bibfield  {title} {\bibinfo {title} {{Topological charge excitations and Green's function zeros in paramagnetic Mott insulators}},\ }\href {https://doi.org/10.1103/4k4y-4hj4} {\bibfield  {journal} {\bibinfo  {journal} {Phys. Rev. B}\ }\textbf {\bibinfo {volume} {112}},\ \bibinfo {pages} {085105} (\bibinfo {year} {2025})}\BibitemShut {NoStop}%
\bibitem [{\citenamefont {Wang}\ \emph {et~al.}(2026)\citenamefont {Wang}, \citenamefont {Liu},\ and\ \citenamefont {Moore}}]{Wang2026}%
  \BibitemOpen
  \bibfield  {author} {\bibinfo {author} {\bibfnamefont {Y.-Q.}\ \bibnamefont {Wang}}, \bibinfo {author} {\bibfnamefont {C.}~\bibnamefont {Liu}},\ and\ \bibinfo {author} {\bibfnamefont {J.~E.}\ \bibnamefont {Moore}},\ }\bibfield  {title} {\bibinfo {title} {{Structure of domain walls in chiral spin liquids}},\ }\href {https://doi.org/10.1073/pnas.2601093123} {\bibfield  {journal} {\bibinfo  {journal} {Proc. Natl. Acad. Sci. U.S.A.}\ }\textbf {\bibinfo {volume} {123}},\ \bibinfo {pages} {e2601093123} (\bibinfo {year} {2026})}\BibitemShut {NoStop}%
\bibitem [{\citenamefont {Savary}\ and\ \citenamefont {Balents}(2017)}]{Savary2016QSL}%
  \BibitemOpen
  \bibfield  {author} {\bibinfo {author} {\bibfnamefont {L.}~\bibnamefont {Savary}}\ and\ \bibinfo {author} {\bibfnamefont {L.}~\bibnamefont {Balents}},\ }\bibfield  {title} {\bibinfo {title} {{Quantum spin liquids: a review}},\ }\href {https://doi.org/10.1088/0034-4885/80/1/016502} {\bibfield  {journal} {\bibinfo  {journal} {Rep. Prog. Phys.}\ }\textbf {\bibinfo {volume} {80}},\ \bibinfo {pages} {016502} (\bibinfo {year} {2017})}\BibitemShut {NoStop}%
\bibitem [{\citenamefont {Wagner}\ \emph {et~al.}(2023)\citenamefont {Wagner}, \citenamefont {Crippa}, \citenamefont {Amaricci}, \citenamefont {Hansmann}, \citenamefont {Klett}, \citenamefont {König}, \citenamefont {Schäfer}, \citenamefont {Di~Sante}, \citenamefont {Cano}, \citenamefont {Millis}, \citenamefont {Georges},\ and\ \citenamefont {Sangiovanni}}]{Wagner2023}%
  \BibitemOpen
  \bibfield  {author} {\bibinfo {author} {\bibfnamefont {N.}~\bibnamefont {Wagner}}, \bibinfo {author} {\bibfnamefont {L.}~\bibnamefont {Crippa}}, \bibinfo {author} {\bibfnamefont {A.}~\bibnamefont {Amaricci}}, \bibinfo {author} {\bibfnamefont {P.}~\bibnamefont {Hansmann}}, \bibinfo {author} {\bibfnamefont {M.}~\bibnamefont {Klett}}, \bibinfo {author} {\bibfnamefont {E.~J.}\ \bibnamefont {König}}, \bibinfo {author} {\bibfnamefont {T.}~\bibnamefont {Schäfer}}, \bibinfo {author} {\bibfnamefont {D.}~\bibnamefont {Di~Sante}}, \bibinfo {author} {\bibfnamefont {J.}~\bibnamefont {Cano}}, \bibinfo {author} {\bibfnamefont {A.~J.}\ \bibnamefont {Millis}}, \bibinfo {author} {\bibfnamefont {A.}~\bibnamefont {Georges}},\ and\ \bibinfo {author} {\bibfnamefont {G.}~\bibnamefont {Sangiovanni}},\ }\bibfield  {title} {\bibinfo {title} {{Mott insulators with boundary zeros}},\ }\href {http://dx.doi.org/10.1038/s41467-023-42773-7} {\bibfield  {journal} {\bibinfo  {journal} {Nat Commun}\ }\textbf {\bibinfo {volume} {14}},\
  \bibinfo {pages} {7531} (\bibinfo {year} {2023})}\BibitemShut {NoStop}%
\bibitem [{\citenamefont {Kuhlenkamp}\ \emph {et~al.}(2024)\citenamefont {Kuhlenkamp}, \citenamefont {Kadow}, \citenamefont {Imamoğlu},\ and\ \citenamefont {Knap}}]{Knap2024}%
  \BibitemOpen
  \bibfield  {author} {\bibinfo {author} {\bibfnamefont {C.}~\bibnamefont {Kuhlenkamp}}, \bibinfo {author} {\bibfnamefont {W.}~\bibnamefont {Kadow}}, \bibinfo {author} {\bibfnamefont {A.}~\bibnamefont {Imamoğlu}},\ and\ \bibinfo {author} {\bibfnamefont {M.}~\bibnamefont {Knap}},\ }\bibfield  {title} {\bibinfo {title} {{Chiral Pseudospin Liquids in Moir\'e Heterostructures}},\ }\href {https://doi.org/10.1103/PhysRevX.14.021013} {\bibfield  {journal} {\bibinfo  {journal} {Phys. Rev. X}\ }\textbf {\bibinfo {volume} {14}},\ \bibinfo {pages} {021013} (\bibinfo {year} {2024})}\BibitemShut {NoStop}%
\bibitem [{\citenamefont {Divic}\ \emph {et~al.}(2026)\citenamefont {Divic}, \citenamefont {Soejima}, \citenamefont {Cr\'epel}, \citenamefont {Zaletel},\ and\ \citenamefont {Millis}}]{Divic2026}%
  \BibitemOpen
  \bibfield  {author} {\bibinfo {author} {\bibfnamefont {S.}~\bibnamefont {Divic}}, \bibinfo {author} {\bibfnamefont {T.}~\bibnamefont {Soejima}}, \bibinfo {author} {\bibfnamefont {V.}~\bibnamefont {Cr\'epel}}, \bibinfo {author} {\bibfnamefont {M.~P.}\ \bibnamefont {Zaletel}},\ and\ \bibinfo {author} {\bibfnamefont {A.}~\bibnamefont {Millis}},\ }\bibfield  {title} {\bibinfo {title} {{Chiral spin liquid and quantum phase transition in the triangular-lattice Hofstadter-Hubbard model}},\ }\href {https://doi.org/10.1103/g84x-qwrk} {\bibfield  {journal} {\bibinfo  {journal} {Phys. Rev. B}\ }\textbf {\bibinfo {volume} {113}},\ \bibinfo {pages} {L121107} (\bibinfo {year} {2026})}\BibitemShut {NoStop}%
\bibitem [{\citenamefont {Gallegos}\ \emph {et~al.}(2026)\citenamefont {Gallegos}, \citenamefont {Magaldi}, \citenamefont {Millis},\ and\ \citenamefont {White}}]{us2026}%
  \BibitemOpen
  \bibfield  {author} {\bibinfo {author} {\bibfnamefont {C.~A.}\ \bibnamefont {Gallegos}}, \bibinfo {author} {\bibfnamefont {R.~M.}\ \bibnamefont {Magaldi}}, \bibinfo {author} {\bibfnamefont {A.}~\bibnamefont {Millis}},\ and\ \bibinfo {author} {\bibfnamefont {S.~R.}\ \bibnamefont {White}},\ }\bibfield  {title} {\bibinfo {title} {{Quantum Hall to chiral spin liquid transition in a triangular lattice Hofstadter-Hubbard model}},\ }\href {https://doi.org/10.1103/xv65-m7m8} {\bibfield  {journal} {\bibinfo  {journal} {Phys. Rev. B}\ }\textbf {\bibinfo {volume} {113}},\ \bibinfo {pages} {064412} (\bibinfo {year} {2026})}\BibitemShut {NoStop}%
\bibitem [{\citenamefont {Kalmeyer}\ and\ \citenamefont {Laughlin}(1987)}]{Kalmeyer87}%
  \BibitemOpen
  \bibfield  {author} {\bibinfo {author} {\bibfnamefont {V.}~\bibnamefont {Kalmeyer}}\ and\ \bibinfo {author} {\bibfnamefont {R.~B.}\ \bibnamefont {Laughlin}},\ }\bibfield  {title} {\bibinfo {title} {{Equivalence of the resonating-valence-bond and fractional quantum Hall states}},\ }\href {https://doi.org/10.1103/PhysRevLett.59.2095} {\bibfield  {journal} {\bibinfo  {journal} {Phys. Rev. Lett.}\ }\textbf {\bibinfo {volume} {59}},\ \bibinfo {pages} {2095} (\bibinfo {year} {1987})}\BibitemShut {NoStop}%
\bibitem [{Note1()}]{Note1}%
  \BibitemOpen
  \bibinfo {note} {The vertical bonds pick up an additional phase of $\Phi /L_y$, while the extra phase of the diagonal bonds is smaller by a factor of two (i.e. $\sin \pi /6$); see Figs.~\ref {Fig:Pump}(a) and (b).}\BibitemShut {Stop}%
\bibitem [{\citenamefont {Fishman}\ and\ \citenamefont {White}(2015)}]{Fishman2015GaussianMPS}%
  \BibitemOpen
  \bibfield  {author} {\bibinfo {author} {\bibfnamefont {M.~T.}\ \bibnamefont {Fishman}}\ and\ \bibinfo {author} {\bibfnamefont {S.~R.}\ \bibnamefont {White}},\ }\bibfield  {title} {\bibinfo {title} {{Compression of correlation matrices and an efficient method for forming matrix product states of fermionic Gaussian states}},\ }\href {https://doi.org/10.1103/PhysRevB.92.075132} {\bibfield  {journal} {\bibinfo  {journal} {Phys. Rev. B}\ }\textbf {\bibinfo {volume} {92}},\ \bibinfo {pages} {075132} (\bibinfo {year} {2015})}\BibitemShut {NoStop}%
\bibitem [{\citenamefont {Zaletel}\ \emph {et~al.}(2014)\citenamefont {Zaletel}, \citenamefont {Mong},\ and\ \citenamefont {Pollmann}}]{Zaletel2014}%
  \BibitemOpen
  \bibfield  {author} {\bibinfo {author} {\bibfnamefont {M.~P.}\ \bibnamefont {Zaletel}}, \bibinfo {author} {\bibfnamefont {R.~S.~K.}\ \bibnamefont {Mong}},\ and\ \bibinfo {author} {\bibfnamefont {F.}~\bibnamefont {Pollmann}},\ }\bibfield  {title} {\bibinfo {title} {{Flux insertion, entanglement, and quantized responses}},\ }\href {https://doi.org/10.1088/1742-5468/2014/10/P10007} {\bibfield  {journal} {\bibinfo  {journal} {J. Stat. Mech.}\ }\textbf {\bibinfo {volume} {2014}},\ \bibinfo {pages} {P10007} (\bibinfo {year} {2014})}\BibitemShut {NoStop}%
\bibitem [{\citenamefont {Knolle}\ and\ \citenamefont {Moessner}(2019)}]{Knolle2019QSL}%
  \BibitemOpen
  \bibfield  {author} {\bibinfo {author} {\bibfnamefont {J.}~\bibnamefont {Knolle}}\ and\ \bibinfo {author} {\bibfnamefont {R.}~\bibnamefont {Moessner}},\ }\bibfield  {title} {\bibinfo {title} {{A Field Guide to Spin Liquids}},\ }\href {https://doi.org/10.1146/annurev-conmatphys-031218-013401} {\bibfield  {journal} {\bibinfo  {journal} {Annu. Rev. Condens. Matter Phys.}\ }\textbf {\bibinfo {volume} {10}},\ \bibinfo {pages} {451} (\bibinfo {year} {2019})}\BibitemShut {NoStop}%
\bibitem [{\citenamefont {Kallin}\ and\ \citenamefont {Halperin}(1984)}]{Halperin1984}%
  \BibitemOpen
  \bibfield  {author} {\bibinfo {author} {\bibfnamefont {C.}~\bibnamefont {Kallin}}\ and\ \bibinfo {author} {\bibfnamefont {B.~I.}\ \bibnamefont {Halperin}},\ }\bibfield  {title} {\bibinfo {title} {{Excitations from a filled Landau level in the two-dimensional electron gas}},\ }\href {https://doi.org/10.1103/PhysRevB.30.5655} {\bibfield  {journal} {\bibinfo  {journal} {Phys. Rev. B}\ }\textbf {\bibinfo {volume} {30}},\ \bibinfo {pages} {5655} (\bibinfo {year} {1984})}\BibitemShut {NoStop}%
\bibitem [{\citenamefont {Ebisu}\ \emph {et~al.}(2020)\citenamefont {Ebisu}, \citenamefont {Kalloor}, \citenamefont {Tsvelik},\ and\ \citenamefont {Oreg}}]{Ebisu2020}%
  \BibitemOpen
  \bibfield  {author} {\bibinfo {author} {\bibfnamefont {H.}~\bibnamefont {Ebisu}}, \bibinfo {author} {\bibfnamefont {R.~R.}\ \bibnamefont {Kalloor}}, \bibinfo {author} {\bibfnamefont {A.~M.}\ \bibnamefont {Tsvelik}},\ and\ \bibinfo {author} {\bibfnamefont {Y.}~\bibnamefont {Oreg}},\ }\bibfield  {title} {\bibinfo {title} {{Chiral topologically ordered insulating phases in arrays of interacting integer quantum Hall islands}},\ }\href {https://doi.org/10.1103/PhysRevB.102.165112} {\bibfield  {journal} {\bibinfo  {journal} {Phys. Rev. B}\ }\textbf {\bibinfo {volume} {102}},\ \bibinfo {pages} {165112} (\bibinfo {year} {2020})}\BibitemShut {NoStop}%
\bibitem [{foo()}]{footnote}%
  \BibitemOpen
  \href@noop {} {}\bibinfo {note} {We thank an anonymous referee for suggesting this edge-theory interpretation.}\BibitemShut {Stop}%
\bibitem [{\citenamefont {Divic}\ \emph {et~al.}(2025)\citenamefont {Divic}, \citenamefont {Crépel}, \citenamefont {Soejima}, \citenamefont {Song}, \citenamefont {Millis}, \citenamefont {Zaletel},\ and\ \citenamefont {Vishwanath}}]{Divic2025}%
  \BibitemOpen
  \bibfield  {author} {\bibinfo {author} {\bibfnamefont {S.}~\bibnamefont {Divic}}, \bibinfo {author} {\bibfnamefont {V.}~\bibnamefont {Crépel}}, \bibinfo {author} {\bibfnamefont {T.}~\bibnamefont {Soejima}}, \bibinfo {author} {\bibfnamefont {X.-Y.}\ \bibnamefont {Song}}, \bibinfo {author} {\bibfnamefont {A.~J.}\ \bibnamefont {Millis}}, \bibinfo {author} {\bibfnamefont {M.~P.}\ \bibnamefont {Zaletel}},\ and\ \bibinfo {author} {\bibfnamefont {A.}~\bibnamefont {Vishwanath}},\ }\bibfield  {title} {\bibinfo {title} {{Anyon superconductivity from topological criticality in a Hofstadter–Hubbard model}},\ }\href {https://doi.org/10.1073/pnas.2426680122} {\bibfield  {journal} {\bibinfo  {journal} {Proc. Natl. Acad. Sci. U.S.A.}\ }\textbf {\bibinfo {volume} {122}},\ \bibinfo {pages} {e2426680122} (\bibinfo {year} {2025})}\BibitemShut {NoStop}%
\bibitem [{\citenamefont {Kuhlenkamp}\ \emph {et~al.}(2025)\citenamefont {Kuhlenkamp}, \citenamefont {Divic}, \citenamefont {Zaletel}, \citenamefont {Soejima},\ and\ \citenamefont {Vishwanath}}]{kuhlenkamp2025}%
  \BibitemOpen
  \bibfield  {author} {\bibinfo {author} {\bibfnamefont {C.}~\bibnamefont {Kuhlenkamp}}, \bibinfo {author} {\bibfnamefont {S.}~\bibnamefont {Divic}}, \bibinfo {author} {\bibfnamefont {M.~P.}\ \bibnamefont {Zaletel}}, \bibinfo {author} {\bibfnamefont {T.}~\bibnamefont {Soejima}},\ and\ \bibinfo {author} {\bibfnamefont {A.}~\bibnamefont {Vishwanath}},\ }\href@noop {} {\bibinfo {title} {{Robust superconductivity upon doping chiral spin liquid and Chern insulators in a Hubbard-Hofstadter model}}} (\bibinfo {year} {2025}),\ \Eprint {https://arxiv.org/abs/2509.02675} {arXiv:2509.02675 [cond-mat.str-el]} \BibitemShut {NoStop}%
\bibitem [{\citenamefont {Chen}\ \emph {et~al.}(2026)\citenamefont {Chen}, \citenamefont {Wang}, \citenamefont {Zhang}, \citenamefont {Balents},\ and\ \citenamefont {Sheng}}]{Chen2026}%
  \BibitemOpen
  \bibfield  {author} {\bibinfo {author} {\bibfnamefont {F.}~\bibnamefont {Chen}}, \bibinfo {author} {\bibfnamefont {W.~O.}\ \bibnamefont {Wang}}, \bibinfo {author} {\bibfnamefont {J.-X.}\ \bibnamefont {Zhang}}, \bibinfo {author} {\bibfnamefont {L.}~\bibnamefont {Balents}},\ and\ \bibinfo {author} {\bibfnamefont {D.~N.}\ \bibnamefont {Sheng}},\ }\bibfield  {title} {\bibinfo {title} {{Topological Chiral Superconductivity in the Triangular-Lattice Hofstadter-Hubbard Model}},\ }\href {https://doi.org/10.1103/tm9q-w5y7} {\bibfield  {journal} {\bibinfo  {journal} {Phys. Rev. Lett.}\ }\textbf {\bibinfo {volume} {136}},\ \bibinfo {pages} {086503} (\bibinfo {year} {2026})}\BibitemShut {NoStop}%
\bibitem [{\citenamefont {Niu}\ \emph {et~al.}(2026)\citenamefont {Niu}, \citenamefont {Sheng},\ and\ \citenamefont {Peng}}]{Niu2026}%
  \BibitemOpen
  \bibfield  {author} {\bibinfo {author} {\bibfnamefont {S.}~\bibnamefont {Niu}}, \bibinfo {author} {\bibfnamefont {D.~N.}\ \bibnamefont {Sheng}},\ and\ \bibinfo {author} {\bibfnamefont {Y.}~\bibnamefont {Peng}},\ }\href@noop {} {\bibinfo {title} {{Thermodynamic-Limit Evidence for Chiral Superconductivity Induced by Doping Chiral Topological Phases}}} (\bibinfo {year} {2026}),\ \Eprint {https://arxiv.org/abs/2512.21503} {arXiv:2512.21503 [cond-mat.str-el]} \BibitemShut {NoStop}%
\bibitem [{\citenamefont {Fishman}\ \emph {et~al.}(2022)\citenamefont {Fishman}, \citenamefont {White},\ and\ \citenamefont {Stoudenmire}}]{itensor}%
  \BibitemOpen
  \bibfield  {author} {\bibinfo {author} {\bibfnamefont {M.}~\bibnamefont {Fishman}}, \bibinfo {author} {\bibfnamefont {S.~R.}\ \bibnamefont {White}},\ and\ \bibinfo {author} {\bibfnamefont {E.~M.}\ \bibnamefont {Stoudenmire}},\ }\bibfield  {title} {\bibinfo {title} {{The ITensor Software Library for Tensor Network Calculations}},\ }\href {https://doi.org/10.21468/SciPostPhysCodeb.4} {\bibfield  {journal} {\bibinfo  {journal} {SciPost Phys. Codebases}\ ,\ \bibinfo {pages} {4}} (\bibinfo {year} {2022})}\BibitemShut {NoStop}%
\end{thebibliography}%

\newpage

\onecolumngrid
\begin{center}
\vskip 0.1cm
{\large\bf End Matter}
\end{center}
\vskip 0.5cm
\twocolumngrid

\renewcommand{\theequation}{A\arabic{equation}}
\setcounter{equation}{0}
{\it Convergence of excited states.}---%
The relaxation of an excited state involves both optimizing the local structure of the quasiparticle and relaxing its position and wavepacket shape.
We find that a finite bond dimension $\chi$ restricts the spatial spreading of the excitation---apparently, there is a cost in bond dimension to having a very broad wavepacket. In the absence of truncation, a wavepacket spreads indefinitely to lower its kinetic energy. With truncation, there is a counteracting tendency, constraining the width. Apparently the states needed to construct a plane wave are not included in the fixed basis implied by the finite bond dimension.
This makes the finite-$\chi$ states useful diagnostics, since the spatial structure in the sweep optimized fixed bond dimension states enable easy identification of characteristic structures of the state, such as spin-charge separation of the single-particle excitation in the CSL.  This subtle interplay is evident in Figures~\ref{Fig:1p}-\ref{Fig:Interface1pSp} in the main text.

\begin{figure}[t!]
\includegraphics[width=.95\linewidth]{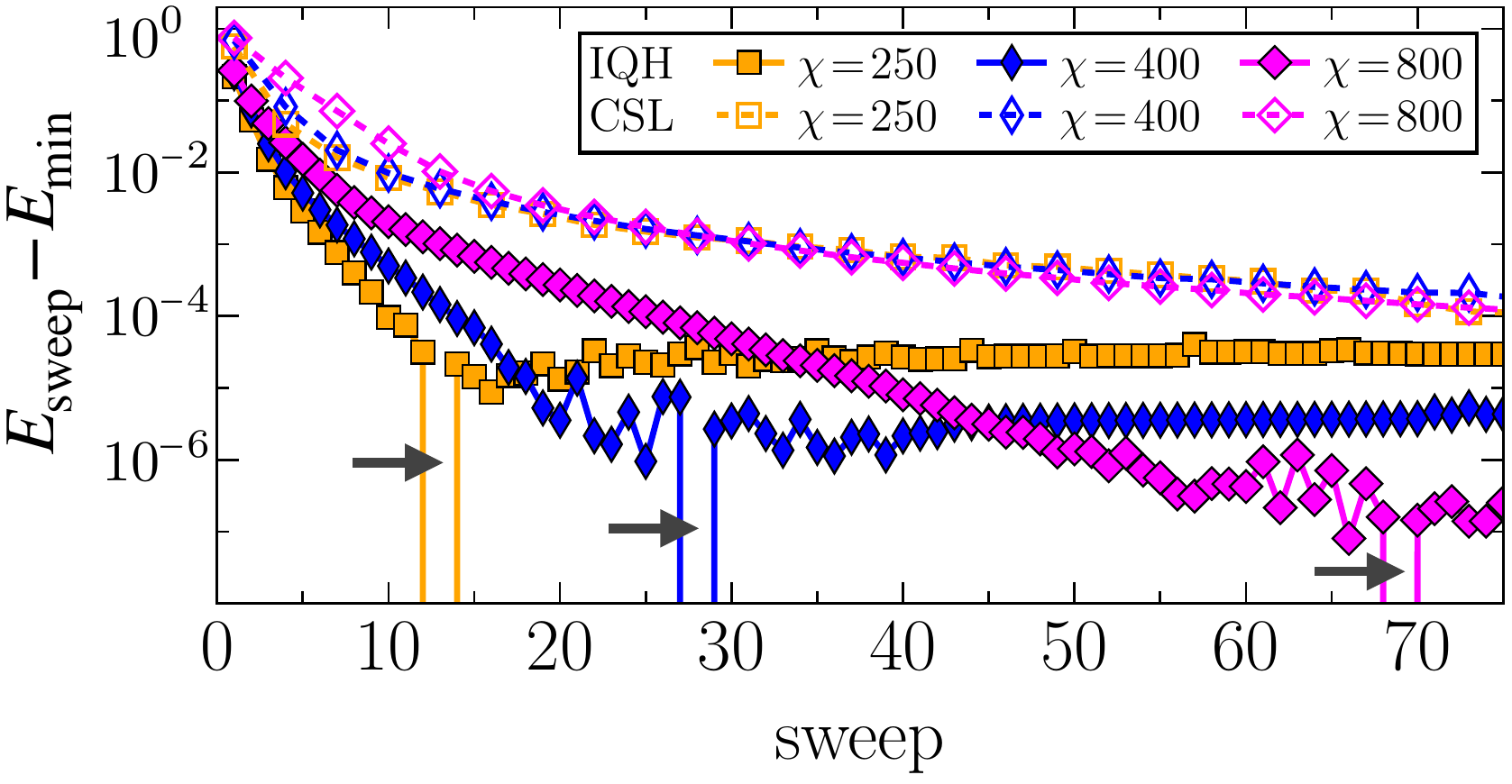}
\vskip -0.3cm
\caption{Energy on each sweep, $E_{\rm sweep}$, relative to the minimum energy $E_{\rm min}$ reached at fixed bond dimension $\chi$ in a $240\times3$ cylinder.
The full symbols with solid lines and the open symbols with dashed lines show the IQH and CSL energies, respectively.
Orange, blue, and magenta correspond to $\chi\!=\!250$, $400$, and $800$, respectively.
The arrows indicate the sweep number at which the minimum energy in the IQH phase is reached.}
\label{Fig:Esweeps}
\vskip -0.5cm
\end{figure}

From an initial unoptimized, small-width wavepacket, the rate at which an excitation lowers its energy (measured in sweeps) varies.
Figure~\ref{Fig:Esweeps} shows the energy on each sweep, $E_{\rm sweep}$, relative to the minimum energy $E_{\rm min}$ reached at fixed bond dimension $\chi$, in a $240\times3$ cylinder.
The full symbols with solid lines and the open symbols with dashed lines show the IQH and CSL energies, respectively.
The orange, blue, and magenta colors correspond to $\chi\!=\!250$, $400$, and $800$, respectively.
The arrows indicate the sweep number at which the minimum energy in the IQH phase is reached.
After the minimum, the energy fluctuates around its lowest value.

In contrast, the CSL energy decreases more slowly.
For $\chi=250$, it decreases monotonically up to about $150$ sweeps, where it reaches its minimum (not shown).
For the larger bond dimensions, $\chi=400$ and $800$, the CSL energy continues to decrease after $200$ sweeps and does not exhibit a minimum within the range studied.
This behavior is consistent with the slow separation of the spinon and chargon, which provides an additional route for lowering the energy.
Thus, the excited states in the CSL should be viewed as diagnostics of the excitation structure rather than as fully converged lowest-energy states in their quantum-number sector.

\begin{figure}[b!]
\includegraphics[width=\linewidth]{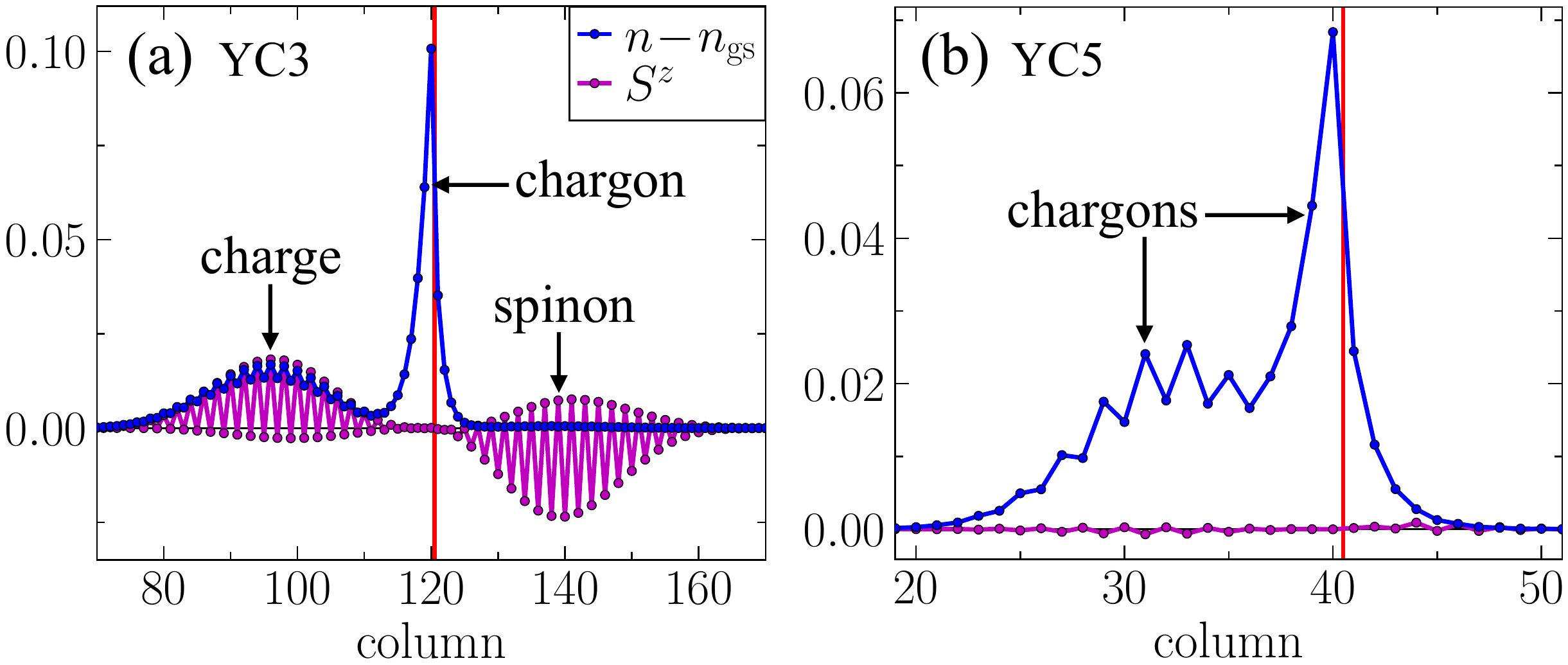}
\vskip -0.1cm
\caption{Two-particle excited states in the IQH--CSL interface in the (a) YC3 and (b) YC5 cylinders. 
The solid red, blue, and purple lines correspond to the IQH-CSL interface, the deviation of the charge density from the ground state, $n\!-\!n_{\rm gs}$, and the local magnetization, $S^z$, respectively. 
The charge-like excitation, the chargons, and the spinon are indicated.}
\label{Fig:Interface2p}
\vskip -0.5cm
\end{figure}

{\it Two-particle excited states.}---%
We further characterize the spinless charged mode at the IQH--CSL interface by calculating two-particle excited states.
Figure~\ref{Fig:Interface2p} shows the states obtained by creating two electrons with opposite spin on the YC3 and YC5 cylinders, Figs.~\ref{Fig:Interface2p}(a) and \ref{Fig:Interface2p}(b), respectively.

For the narrower YC3 cylinder, Fig.~\ref{Fig:Interface2p}(a), one of the electrons fractionalizes. The chargon remains bound to the interface, while the spinon is weakly repelled into the CSL bulk, in close analogy with the single-particle excitation discussed in the main text; see Fig.~\ref{Fig:Interface1pSp}(a).
The second electron remains close to the interface, but is displaced toward the IQH side.

The behavior on the wider YC5 cylinder, Fig.~\ref{Fig:Interface2p}(b), is qualitatively different.
In this case we observe only spinless charge excitations in the relaxed density profiles.
This suggests a process in which the two electrons fractionalize, while the spin densities of the two spinons, with opposite spin $S^z=\pm 1/2$, cancel in the total $S^z$ profile.
At intermediate bond dimension, both chargons are tightly bound to the interface.
However, upon increasing the bond dimension to $\chi=6000$, one chargon is pushed away from the interface and into the IQH side.

Taken together, these results suggest that the spinless charged excitations at the IQH--CSL interface are repulsively interacting.
This effective repulsion is most apparent on the YC3 cylinder, whose narrow width cannot accommodate two chargons at the interface.


\end{document}